\documentclass[usenatbib, hyphens]{mnras}

\usepackage{amssymb}
\usepackage{amsmath}
\usepackage{natbib}
\usepackage[pdftex]{graphicx}
\usepackage[normalem]{ulem}
\usepackage{mathtools}
\usepackage{cuted}

\defcitealias{murga16a}{Paper~I}

\title[Impact of HAC evolution on the formation of small hydrocarbons]
{Impact of HAC evolution on the formation of small hydrocarbons in the Orion Bar and the Horsehead PDRs}
\author[M. S. Murga et al.]
{M. S. Murga$^{1,2}$\thanks{E-mail: murga@inasan.ru},
 A. I. Vasyunin$^{2}$, M. S. Kirsanova$^{1}$\\
$^{1}$Institute of Astronomy, Russian Academy of Sciences, Pyatnitskaya str. 48, Moscow 119017, Russia \\
$^{2}$ Institute of Natural Sciences and Mathematics, Ural Federal University,
19 Mira Str., 620075 Ekaterinburg, Russia}
\date{Accepted today. Received tomorrow; in original form \today}
\pubyear{2021}

\begin{document}
\label{firstpage}
\pagerange{\pageref{firstpage}--\pageref{lastpage}}\maketitle

\begin{abstract}
We study evolution of hydrogenated amorphous carbon (HAC) grains under harsh UV radiation in photo-dissociation regions (PDRs) near young massive stars. Our aim is to evaluate the impact of the HAC grains on formation of observed small hydrocarbons: C$_2$H, C$_2$H$_2$, C$_3$H$^+$, C$_3$H, C$_3$H$_2$, C$_4$H, in PDRs. We developed a microscopic model of the HAC grains based on available experimental results. The model includes processes of photo- and thermodesorption, accretion of hydrogen and carbon atoms and subsequent formation of carbonaceous mantle on dust surface. H$_2$, CH$_4$, C$_2$H$_2$, C$_2$H$_4$, C$_2$H$_6$,  C$_3$H$_4$,  C$_3$H$_6$, C$_3$H$_8$ are considered as the main fragments of the HAC photo-destruction. We simulated evolution of the HAC grains under the physical conditions of two PDRs, the Orion Bar and the Horsehead nebula. We estimated the production rates of the HAC’ fragments in gas phase chemical reactions and compared them with the production rates of fragments due to the HAC destruction. The latter rates may dominate under some conditions, namely, at $A_{\rm V}=0.1$ in both PDRs. We coupled our model with the gas-grain chemical model {\tt MONACO} and calculated abundances of observed small hydrocarbons. We conclude that the contribution of the HAC destruction fragments to chemistry is not enough to match the observed abundances, although it increases the abundances by several orders of magnitude in the Orion Bar at $A_{\rm V}=0.1$. Additionally, we found that the process of carbonaceous mantle formation on dust surface can be an inhibitor for the formation of observed small hydrocarbons in PDRs.
\end{abstract}


\section{Introduction}

Carbon is one of the most abundant elements in the interstellar medium (ISM). It can exist either in atomic form or locked in a mixed or pure carbonaceous compound. These compounds may contain from a few atoms to several thousands of carbon atoms. Vast variety of the carbonaceous compounds and their sensitivity to outer conditions is a topic of intensive studies. One of the unresolved questions is the relatively high abundance of small hydrocarbons (CH$_4$, C$_3$H$_2$, C$_2$H, etc.) in photo-dissociation regions (PDRs) which is not explained by models with gas phase chemical reactions~\citep{pety05, guzman15}.

As carbon is locked in dust grains, and these dust grains can undergo photo-destruction under harsh ultraviolet radiation~\citep{murga19}, it is quite logical to suppose the `top-down' scenario according to which some small hydrocarbons can form due to the photo-destruction of the dust grains. This scenario was also indirectly supported by the observations in the Horsehead nebula where the peak of the emission at 8~$\mu$m which is believed to arise from vibrations of some aromatic hydrocarbons coincides with the peak of the emission of small hydrocarbons~\citep{pety05}. In our previous work by \cite{murga20_acet} we considered photo-dissociation of polycyclic aromatic hydrocarbons (PAHs) and their contribution to the abundance of small hydrocarbons in two PDRs, the Orion Bar and the Horsehead nebula. We showed that taking into account of the PAH dissociation does not increase modelled abundances up to the observational values. However, PAHs represent are just one of the types of the interstellar carbonaceous dust while other forms exist along with them. According to the model presented in the work of \cite{jones13} the material of carbonaceous dust is rather hydrogenated amorphous carbon (HAC) which becomes dehydrogenated in the ISM due to influence of the ultraviolet (UV) radiation~\citep{welch72, gruzdkov94, mennella01}. The experiments on photo-desorption of HAC filaments are promising: they indicate that rates of production of fragments are relatively high and fragments can be different (CH$_4$, C$_2$H$_2$, C$_3$H$_4$, etc.)\footnote{Note that fragments of HAC destruction are also small hydrocarbons, however we use the term `fragments' if these small hydrocarbons are products of HAC destruction and the term `small hydrocarbons' if observed small hydrocarbons are meant even though they are the same.} (e.g. \cite{alata14, alata15, duley15}) unlike PAHs which dissociation leads mostly to formation of C$_2/$C$_2$H$_2$ molecules~\citep{jochims94}. In addition, based on the experiments of \cite{alata14, alata15} where thickness of HAC filaments is around 1--5$\mu$m, we suppose that large HAC grains can be photo-processed (at least their surface) while only small PAHs ($N_{\rm C}\lesssim50$) can be dissociated and lose hydrocarbons~\citep{allain96, zhen15}. Thus, a larger number of grains can potentially provide hydrocarbons if to consider HAC grains instead of PAHs. Recently \cite{awad22} estimated the contribution of fragments of the HAC destruction into the chemical composition of diffuse clouds. They concluded that the inclusion of the HAC destruction products into chemical network of reactions could explain the observations of small hydrocarbons. Nevertheless, the model in the work of \cite{awad22} does not include the microscopic treatment with detailed modelling of individual grain evolution. It is not considered that HAC grains may be quickly dehydrogenated in the ISM and, correspondingly, be not able to produce the hydrocarbons obtained in the experiment with highly hydrogenated HAC films. According to \cite{jones13} HAC grains are rather dehydrogenated in the ISM, and only large grains can be still H-rich but  inside a grain, i.e. they can not produce hydrocarbons. Thus, the goal of this work is to model the HAC evolution with the microscopic treatment and to estimate their contribution to the abundance of small hydrocarbons.

In this work, we develop the model of the HAC evolution under conditions of PDRs. We adapt this model to the same PDRs, the Orion Bar and the Horsehead nebula, which were considered in our previous work by \cite{murga20_acet}. We implement the results of the HAC evolution model to the chemical model of gas phase chemical reactions and estimate abundances of observed hydrocarbons.

\section{HAC evolution model}

\subsection{Kinetic equations}

\begin{figure}
\includegraphics[width=0.45\textwidth]{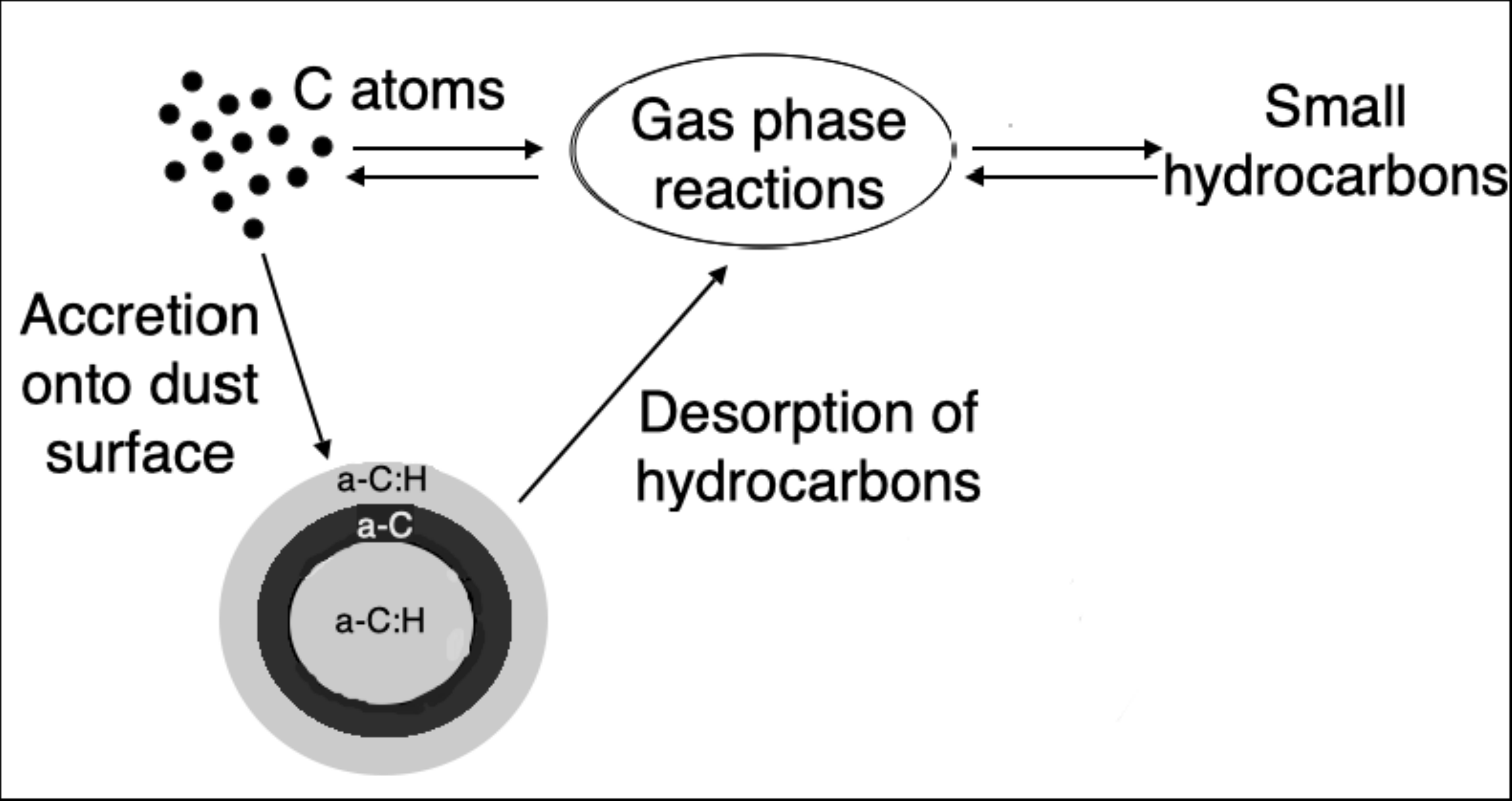}
\caption{The lifecycle of carbons atoms in our model.}
\label{fig: scheme}
\end{figure}

The general scheme of our model is shown in Fig.~\ref{fig: scheme} which also illustrates the adopted scenario of the lifecycle of carbon atoms. Carbon atoms are distributed between the gas phase species and carbonaceous dust grains. Gaseous carbon may participate in gas phase reactions including the reactions of formation of different small hydrocarbons. Along with it carbon may accrete onto the dust surface and form the carbonaceous amorphous mantle. This mantle is formed via chemisorption and hard relative to an icy mantle which is formed in molecular clouds. Carbonaceous fragments desorbed from the grain surface also participate in gas phase reactions. Small hydrocarbons can be dissociated due to photolysis, and, therefore, carbon atoms release back to gas. Our model includes the concomitant processes of this scenario: photo- and thermo-desorption of different fragments from the HAC grains, accretion of hydrogen and carbons atoms and formation of the carbonaceous amorphous mantle on dust surface due to the accretion. Additionally, we calculate gas phase reactions using the model {\tt MONACO}~\citep{vasyunin13, vasyunin17} which was developed previously and will be described shortly below.  

Kinetic equations for evolution of the HAC grains were drawn up similar to those in the {\tt Shiva} model~\citep{murga19}, although the system of equations was extended. We supplemented the model with processes of thermal desorption of amorphous carbonaceous mantle, accretion of carbon and hydrogen atoms. Equations for gas phase atomic hydrogen, atomic and ionized carbon densities ($n({\rm H})$, $n({\rm C})$, and $n({\rm C}^{+})$ respectively) were added to the system as these densities change due to accretion. In this work we consider several products of dust photo-destruction, the corresponding kinetic equation was added for each of them. 

To describe the evolution of the HAC grains quantitatively, we distribute the HAC grains in the mass range from $m_{\rm min}$ to $m_{\rm max}$ into $N_{\rm m}$ bins by their carbon mass with $N_{\rm m}+1$ borders, $m_{\rm b}^i$. A grain from an $i$th bin has mass $m_i$ and the number of carbon atoms $N_{i}^{\rm C}$. Grains inside each $i$th bin are divided by their hydrogen mass into $N_{\rm mH}$ bins with $N_{\rm mH}+1$ borders. These bins are designated by indices $j$, and their mass borders are $m_{\rm bH}^{ij}$. The number of hydrogen atoms in an $ij$ bin is $N_{ij}^{\rm H}$. Each $ij$th bin is characterised by its number density $n_{ij}$.

Grains can transit from bin to bin during evolution changing the carbon and hydrogen mass. We describe all transitions in the way when only one index changes, i.e. we consider only transitions from $ij$th bin to $(i-1)j$,$(i+1)j$ or to $i(j-1)$,$i(j+1)$ bins (a so-called `five point diffusion scheme'). If a grain changes both carbon and hydrogen mass simultaneously, we divide the transition into two separate transitions with change of only $i$ or $j$. In the case of large mass bins neglecting corner transitions has a small effect on results while significantly speeding up calculations.

During dissociation HAC grains produce $N_{\rm frag}$ different fragment species. The number density of the produced $k$th fragment is designated as $n_{k}$. The numbers of carbon and hydrogen atoms in the $k$th fragment are designated as $N_{k}^{\rm C}$ and $N_{k}^{\rm H}$, respectively. 

The system of equations is as follows:
\newpage
\begin{strip}
\begin{equation} \label{eq: big formula}
\begin{split}
\begin{cases}
\frac{dn_{ij}}{dt} &= \underbrace{A_{ij+1}^{(1)}n_{ij+1} - A_{ij}^{(1)}n_{ij}}_{\text{H loss due to photo-destruction}} + \underbrace{B_{i+1j}^{(1)}n_{i+1j} - B_{ij}^{(1)}n_{ij}}_{\text{C loss due to photo-destruction}}  \\
&+ \underbrace{A^{(2)}_{ij+1}n_{ij+1} - A^{(2)}_{ij}n_{ij}}_{\text{H loss due to thermal desorption}} + \underbrace{B^{(2)}_{i+1j}n_{i+1j} - B^{(2)}_{ij}n_{ij}}_{\text{C loss due to thermal desorption}}  \\
&+\underbrace{A^{(3)}_{ij-1}n({\rm H})n_{ij-1} - A^{(3)}_{ij}n({\rm H})n_{ij}}_{\text{H addtion}}+\underbrace{B^{(3)}_{i-1j}n({\rm C})n_{i-1j}-B_{ij}^{(3)}n({\rm C})n_{ij}}_{\text{C addition}} +\underbrace{B^{(3+)}_{i-1j}n({\rm C}^{+})n_{i-1j}-B_{ij}^{(3+)}n({\rm C}^{+})n_{ij}}_{\text{C$^{+}$ addition}} \\
\frac{dn_k}{dt} &= \sum\limits_{i=1}^{N_{m}}\sum\limits_{j=1}^{N_{m{\rm H}}} \left[B^{(1)\star}_{ijk}n_{ij}-B^{(1)\star}_{i+1jk}n_{i+1j}+ B^{(2)\star}_{ijk}n_{ij}-B^{(2)\star}_{i+1jk}n_{i+1j}\right] \frac{N_{i}^{\rm C} }{N_{k}^{\rm C}}\\
\frac{dn({\rm C})}{dt} &= -\sum\limits_{i=1}^{N_{\rm m}}\sum\limits_{j=1}^{N_{\rm mH}} \left[B^{(3)}_{i-1j}n_{i-1j} - B^{(3)}_{ij}n_{ij}\right] n({\rm C}) N^{\rm C}_{i}\\
\frac{dn({\rm C}^{+})}{dt} &= -\sum\limits_{i=i}^{N_{\rm m}}\sum\limits_{j=1}^{N_{\rm mH}} \left[B^{(3+)}_{i-1j}n_{i-1j} - B^{(3+)}_{ij}n_{ij}\right] n({\rm C^{+}}) N^{\rm C}_{i}\\
\frac{dn({\rm H})}{dt} &= -\sum\limits_{i=1}^{N_{\rm m}}\sum\limits_{j=1}^{N_{\rm mH}} \left[A^{(3)}_{ij-1}n_{ij-1} - A_{ij}^{(3)}n_{ij}\right] n({\rm H})N^{\rm H}_{ij},
\end{cases}
\end{split}
\end{equation}
\end{strip} 
where $A^{(1,2,3)}_{ij}$ and $B^{(1,2,3,3+)}_{ij}$ are the rate coefficients of the hydrogen and carbon mass loss/addition, respectively. The indices 1 and 2 correspond to processes of photo-destruction and thermal desorption, respectively, the index 3 corresponds to the accretion process. Note that units of $A^{(1,2)}_{ij}$ and $B^{(1,2)}_{ij}$ are s$^{-1}$, while units of $A^{(3)}_{ij}$ and $B^{(3,3+)}_{ij}$ are s$^{-1}$~cm$^{3}$. The rate coefficients describe the rate of transition of a grain between neighbouring bins, e.g. from $i(j+1)$th to $ij$th in the case of the H loss or from $(i-1)j$th to $ij$th in the case of the C addition. These coefficients are expressed as 
\begin{eqnarray}
A^{(1,2,3)}_{ij} = \frac{\varepsilon^{(1,2,3)}_{ij}}{m_{{\rm Hb}}^{ij+1}-m_{{\rm Hb}}^{ij}},\\ \nonumber
B^{(1,2,3,3+)}_{ij} = \frac{\mu^{(1,2,3,3+)}_{ij}}{m_{\rm b}^{i+1}-m_{\rm b}^{i}},
\end{eqnarray}
where $\varepsilon^{(1,2,3)}_{ij}$ and $\mu^{(1,2,3,3+)}_{ij}$ are hydrogen and carbon decrements, respectively. The coefficients $B^{(1,2)}_{ij}$ describe the mass loss due to detachment of all carbonaceous fragments altogether, while the mass loss due to detachment of the $k$th fragment species is characterised by $B^{(1, 2)\star}_{ijk}$ which can be expressed as 
\begin{equation}
B^{(1, 2)\star}_{ijk} = \frac{\mu^{(1,2)}_{ijk}}{m_{\rm b}^{i+1}-m_{\rm b}^{i}},
\end{equation}
i.e. $B^{(1,2)}_{ij}=\sum\limits_{k=i}^{N_{\rm frag}}B^{(1, 2)\star}_{ijk}$. The mass decrements are expressed as
\begin{equation}
\varepsilon^{(1)}_{ij} = \frac{\mu_{\rm H}}{N_{\rm A}}\sum\limits_{k=1}^{N_{\rm frag}}N^{\rm H}_{k}R_{ijk}, 
\end{equation}
\begin{equation}
\varepsilon^{(2)}_{ij} = \frac{\mu_{\rm H}}{N_{\rm A}}\sum\limits_{k=1}^{N_{\rm frag}}N^{\rm H}_{k}T_{ijk}, 
\end{equation}
\begin{equation}
\varepsilon^{(3)}_{ij}  = \frac{\mu_{\rm H}}{N_{\rm A}}k^{\rm H}_{ij},
\end{equation}
\begin{equation}
\mu^{(1)}_{ij} = \frac{\mu_{\rm C}}{N_{\rm A}}\sum\limits_{k=1}^{N_{\rm frag}}N_{k}^{\rm C}R_{ijk}, \;\mu^{(1)}_{ijk} = \frac{\mu_{\rm C}}{N_{\rm A}}N_{k}^{\rm C}R_{ijk},
\end{equation}
\begin{equation}
\mu^{(2)}_{ij} = \frac{\mu_{\rm C}}{N_{\rm A}}\sum\limits_{k=1}^{N_{\rm frag}}N_{k}^{\rm C}T_{ijk}, \; \mu^{(2)}_{ijk} = \frac{\mu_{\rm C}}{N_{\rm A}}N_{k}^{\rm C}T_{ijk},
\end{equation}
\begin{equation}
\mu^{(3,3+)}_{ij} = \frac{\mu_{\rm C}}{N_{\rm A}}k^{\rm C, C^+}_{ij},
\end{equation}
where  $R_{ijk}$ and $T_{ijk}$ are rates of detachment of a $k$th fragment from an $ij$th grain due to photo-destruction and thermal desorption, respectively, $\mu_{\rm H}$ and $\mu_{\rm C}$ are molar masses of hydrogen and carbons atoms, correspondingly, and $N_{\rm A}$ is the Avogadro number, $k^{\rm H}_{ij}$ and  $k^{\rm C, C^+}_{ij}$ are the rates of attachment of hydrogen and carbon atoms due to accretion, correspondingly. The expressions for the rates $R_{ijk}$, $T_{ijk}$, $k^{\rm H}_{ij}$ and $k^{\rm C, C+}_{ij}$ will be given below. The boundary conditions of system~\ref{eq: big formula} are following
\begin{equation} \label{eq: boundaries}
\begin{split}
\begin{cases}
A^{(1,2,3)}_{ij} &= 0 {\text{ if }} j=N_{m{\rm H}} {\text{ or }} j=0 \\
B^{(1,2)}_{ij} &= 0 {\text{ if }} i>N_{m} \\
B^{(3,3+)}_{ij} &= 0 {\text{ if }} i=N_{m} {\text{ or }} i=0 
\end{cases}
\end{split}
\end{equation}

All the rates are sensitive to the grain charge. We calculate the mean charge of grains designated as $Z_{ij}$ following \cite{wd01_ion}, and the evolutionary processes proceed with the rates corresponding to the mean charge.

\subsection{Rates of photo-destruction}

In order to calculate the rates of photo-detachment of a $k$th fragment from an $ij$th grain we apply the formula similar to the one that is used for calculations of the dehydrogenation rate~\citep{jones13, murga19}:
\begin{equation} \label{eq: rijk}
R_{ijk} = 
\left\{
\begin{array}{ll}
Y_{\rm diss}^{\rm k} \sigma^{k}_{\rm loss} N_{ij}^{\rm H} \int\limits_{6.8{\rm eV}}^{13.6{\rm eV}}
Q_{\rm abs}(i,j, Z_{ij}, E) \frac{F(E)}{E} dE, \mbox{if}\; j>0\\
R_{i}({\rm PAH}), \mbox{if}\; j=0
\end{array} \right .
\end{equation}

where $Y_{\rm diss}^{\rm k}$ and  $\sigma^{k}_{\rm loss}$ are the yield and  cross-section of detachment of the $k$th fragment, respectively, $Q_{\rm abs}(i,j, Z_{ij}, E)$ is the absorption efficiency of the $ij$th grain with charge $Z_{ij}$ of photons with energy $E$, and $F(E)$ is the radiation flux which is adopted to be the average interstellar radiation field flux in the Solar neighbourhood estimated in \cite{mmp83} scaled by a factor $\chi$. The list of fragments, their yields and cross-sections are taken from \cite{alata14, alata15}. The value of $\sigma^{k}_{\rm loss}$ is given per one C-H bond, so we multiply it by the number of hydrogen atoms, $N_{ij}^{\rm H}$, which we adopt as the number of C-H bonds in $ij$th grain. The total number of fragments, $N_{\rm frag}$, is 8. Namely, they are H$_2$, CH$_4$, C$_2$H$_2$, C$_2$H$_4$, C$_2$H$_6$, C$_3$H$_4$, C$_3$H$_6$, C$_3$H$_8$. In the experiments of \cite{alata14, alata15}, the UV lamp with photon energies in the range of 6.8-10.5~eV was used. We extend the energy range up to 13.6~eV assuming that photons with energies higher than 10.5~eV influence on the grains in the same way as photons with lower energies. 

According to the {\tt SHIVA} model dehydrogenated HAC grains can lose only C$_2$ atoms. In this case, the destruction rate can be calculated using the treatment that is suitable for PAHs (see details in \cite{murga19}). We estimate this rate designated as $R_{i}({\rm PAH})$ in same way as in \cite{murga20_acet}. The number of atoms in HAC grains is quite large, and the destruction rate is significantly reduced for such large systems. So dehydrogenated HACs practically do not undergo the destruction. 

\subsection{Rates of accretion}

Atoms from gas accrete on dust surface. We are mostly interested in carbon and hydrogen atoms because this accretion may lead to the formation of a HAC mantle via chemisorption~\citep{menela02, jones14, chiar13, jones16}. \cite{jones16} estimated that PDRs have suitable conditions for the mantle formation at $A_{\rm V}\sim0.5-2$. Accordingly, this process was included in our model. It was shown that the hydrogenation process of HAC material is efficient, and the formation of chemical C-H bonds occurs almost without a barrier, herewith the C-C aromatic bonds can be broken for the formation of C-H bond~\citep{menela06, jariwala09}. However, this process is inefficient in molecular clouds where hydrogen is mostly in the molecular phase~\citep{menela06}.  

In fact, the mantle formation process is similar to the deposition of elements on surface which is widely used in engineering with different final aims~\citep{dearnaley05, deposition_book}. A thin coating is formed on the surface as a result of the deposition, and the coating has some specific features like hardness, low friction, inertness, etc. The carbonaceous coating (namely, a-C:H) is a popular one. Due to the importance of these coatings, a plenty of the formation methods have been developed so far~\citep{dearnaley05}. The efficiency of the methods vary significantly, and none of the methods is close to astrophysical conditions, although it can be supposed that carbon and hydrogen attachment under PDR conditions is similar to the ion beam bombardment. It is hard to estimate the efficiency of such astrophysical deposition process. Therefore, we define a coefficient that shows the efficiency of atom incorporation to an existing structure and subsequent formation of a mantle. \cite{jones14} considered a coefficient $\xi$ for the efficiency of incorporation of H atoms to a carbonaceous structure, and this coefficient was varied from 0.01 to 1 in that work. We follow his designation in this work, but we mean that  $\xi$ characterises both C and H incorporation and mantle formation. We vary $\xi$ in the range from 0 to 1 which is slightly wider than the range in \cite{jones14} to demonstrate the case without accretion.

The accretion or attachment rate of an $x$ element on the surface of a grain with radius $a_{i}$ can be calculated as  
\begin{equation} \label{accr}
k_{ij}^{x} =  0.25 \xi F_{\rm C} v_{x} S_{x} 4 \pi a_i^2,
\end{equation}
where $F_{\rm C}$ is the Coulomb factor, $v_{x}$ is a thermal velocity, $S_{x}$ is a sticking coefficient for $x$ element which can be C, C$^{+}$ or H atom, and $a_i$ is a grain radius. According to \cite{jones14} we adopt $S_{{\rm C/C}^{+}}=1$ and $S_{\rm H}=0.3$. The Coulomb factor is calculated by
\begin{equation} \label{coulomb}
F_{\rm C} = 1 - \frac{2 Z_{ij}Z_{x} e^2}{m_{x}a_i v_{x}^2},
\end{equation}
where $Z_x$ and $m_{x}$ are the charge number and the mass of an $x$ element, and $e$ is the electron charge.

The main difference between the carbonaceous mantle we consider here and the icy mantle traditionally considered in models of molecular clouds is the formation of chemical bonds instead of  physical bonds. The forming bonds are aliphatic, and they are believed to be weak relative to strong aromatic bonds, however, they are much stronger than the physical bonds between elements in ice. Binding energies of chemical bonds are about several eV, while binding energies of physical bonds are only a few tenths of eV. That is why the carbonaceous mantle supposedly may exist in the PDR environment where the ice mantle is quickly evaporated due to thermal desorption. Nonetheless, the thermal desorption should be taken into account for the carbonaceous mantle as well, and we describe it below. We note that at some point the ice mantle formation becomes efficient, and it is not correct to consider simultaneously the formation of both mantles because all the elements will be mixed. Using the MONACO gas-grain chemical model, we estimated values of A$_{\rm V}$ at which icy mantles appear on grains in our models of the Orion Bar and Horsehead nebula PDRs. We found that in case of the Horsehead nebula at least one monolayer of water ice on grains is formed at A$_{\rm V}\approx3$. In case of the Orion Bar, icy mantle is not built up until A$_{\rm V}=5$. This is due to much higher UV field intensity in the Orion Bar in comparison to the Horsehead nebula ($\chi=1.2\times10^{4}$ vs. 4.5$\times 10^1$ at A$_{\rm V}=0.1$). Thus, in this work we explore the impact of HAC destruction on chemistry of PDRs with our model up to A$_{\rm V}$=3 in case of the Horsehead nebula and up to A$_{\rm V}$=5 in case of the Orion Bar. We understand that exact values of the threshold optical depth for ice formation are somewhat model dependent and may be different in other studies (e.g. \cite{rollig22}). However, the major impact of HAC destruction on chemistry of PDRs is expected at low visual extinctions. Thus, we believe that for the purpose of this study one does not need to estimate thresholds for ice formation with high accuracy. 

\subsection{Rates of thermal desorption}

We consider hydrogenated carbonaceous mantle which has chemical bonds with dissociation energies higher than physical bonds in the icy mantle as mentioned above. However, dust temperature in PDRs is higher than in molecular clouds, and, consequently, thermal desorption must be quite efficient. Process of the thermal desorption of HAC were studied in laboratory~\citep{wild87, salancon08, peter12, akasaka19}. It is found that the threshold temperature at which HAC material starts to desorb depends on its hydrogenation level and hardness and varies from 300 to 1000~K. At temperature around the threshold only hydrogen atoms desorb, while as the temperature increases carbonaceous fragments start to desorb as well. Difference in the threshold temperatures reflects the difference in the binding energies of hydrogen and carbonaceous species. 

To calculate the rates of the thermal desorption we use the expression for the thermal desorption of icy mantle~\citep{hasegawa92, walch10}. Additionally we take into account that small dust grains (radius is smaller than $\approx$200~\AA{}) have temperature $T_{{\rm d},i}$ in a wide range from $T_{\rm d}^{\rm min}$ to $T_{\rm d}^{\rm max}$ with the probability $\frac{dp}{dT_{{\rm d},i}}$, so the rates should be summed over the range of the temperatures. For large grains the probability function turns to the delta-function at the equilibrium temperature. The expression for the rate is
\begin{equation} \label{desrate}
T_{ijk} = Y_{\rm therm}^{\rm k}\nu\int\limits_{T_{{\rm d}}^{\rm min}}^{T_{{\rm d}}^{\rm max}}\exp\left(\frac{-E_0^{k}}{T_{{\rm d},i}}\right)\frac{dp}{dT_{{\rm d},i}} dT_{{\rm d},i},
\end{equation} 
where $Y_{\rm therm}^{\rm k}$ is the fraction of desorption for the specific $k$th fragment, $\nu$ is a pre-exponential factor, and $E_0^k$ is the binding energy for the $k$th fragment in K. We adopt $T_{{\rm d}}^{\rm min}=1$~K and $T_{{\rm d}}^{\rm max}=5000$~K although the range is typically much narrower. We calculate the temperature probability distribution function following \cite{guhathakurta89} as well as in previous our works~\citep{murga19, murga20_acet}. 

Some estimates of $E_0^{k}$ for hydrogen (i.e. $k=1$) were given in \cite{jacob20} based on the results of their experiment for the `hard' a-C:H material which contains about 30\% of hydrogen. This `hard' material is rather `soft' for the ISM where the fraction of hydrogen can drop to $\sim$0.05~\citep{jones12_1, jones12_2, jones12_3}. We adopt their value of $\nu$ equalled to $10^{16}$~s$^{-1}$ and their mean binding energy for H atoms equalled to $\sim$3.1~eV. They provide information that around 85\% of hydrogen was released with H$_2$, while the other 15\% with different hydrocarbons. We suppose that this proportion is the proportion between the rates of release of corresponding fragments. Juxtaposing the rates obtained with eq.~\ref{desrate} we can roughly estimate the mean value of $E_0^{k}$ ($k=2,...,8$) for hydrocarbons. This value is found to be 3.37~eV. We use this obtained value for all carbonaceous fragments. Judging by the results of \cite{salancon08, peter12} carbonaceous fragments detected after the thermal desorption are roughly the same as after photo-desorption, and the proportions between their signals on mass spectrometer also look pretty similar. As there are no precise measurements of yields of different fragments for the thermal desorption, we adopt that HAC grains lose the same fragments in the same quantitative proportions as in the photo-desorption process. Thus, we find fractions of desorption for the $k$th fragment as $Y_{\rm therm}^{\rm k}=Y_{\rm diss}^{\rm k}/\sum\limits_{k=2}^{8}Y_{\rm diss}^{\rm k}$.

\subsection{Chemical model}

We couple the HAC evolution model with the chemical model MONACO~\citep{vasyunin13, vasyunin17} to estimate the abundance of small hydrocarbons. In our previous work~\citep{murga20_acet} we performed post-processing calculations. However, in this work the calculation scheme was changed and made iterative due to significant loss of gas-phase  carbon (C, C$^{+}$) in the accretion process on the HAC grains. We transfer $n_{\rm C}$, $n_{\rm C^{+}}$, $n_{\rm H}$, $n_{k}$ to the MONACO model as input parameters at each time step. Next, abundances of those species changed due to chemical evolution calculated within the MONACO model, are translated back to the HAC evolution model. Coupling is done in a self-consistent way with mass conservation laws obeyed. The total carbon abundance in the dust and gas phases is conserved, while the exact numbers in the phases change with time.

\subsection{Input parameters for the model}

We adopt the same physical parameters of PDRs ($n_{\rm H}$, the electron number density $n_{\rm e}$, molecular hydrogen density $n_{{\rm H}_2}$, $n_{\rm C}$, $n_{{\rm C}^{+}}$, the gas temperature $T_{\rm gas}$, and $\chi$) as previously in \cite{murga20_acet} (Fig.~2 therein). For the Orion Bar the parameters are taken from \cite{goicoechea15}, and for the Horsehead nebula are from \cite{goicoechea09} ($n_{\rm H}$, $n_{\rm e}$, $n_{\rm C^{+}}$, $n_{\rm C}$) and \cite{legal17} ($n_{\rm H_2}$, $T_{\rm gas}$). The parameters are provided as a function of visual extinction, $A_{\rm V}$. All chemical elements initially are in the atomic form (except hydrogen which can be in the molecular form). Their abundances are standard "low metals"~\citep{wakelam08}. The age of the Orion Bar is adopted to be 10$^5$~yr~\citep{salgado16}, the age of  the Horsehead nebula is 5$\cdot10^{5}$~yr~\citep{pound03}.

The initial number density of dust grains corresponds to the mass distribution presented in \cite{jones13}. The mass borders, $m_{\rm min}$ and $m_{\rm max}$, correspond to radii of 4$\cdot 10^{-8}$ and 4.9$\cdot 10^{-4}$~cm, respectively, according to this distribution. The material density is 1.5~g~cm$^{-3}$. The mass range is divided into $N_{\rm m}=20$ bins. The initial carbon abundance in the dust phase is $1.8\times 10^{-4}$ relative to hydrogen according to \cite{jones13}, which is in 1.3 and 1.8 times higher than the abundances in the gas phase in the Orion Bar adopted from \cite{goicoechea15} and in the Horsehead nebula adopted from \cite{goicoechea09}, correspondingly. The model of \cite{jones13} also involves silicate grains with carbonaceous mantle on it. The mantle is analogous to the one on large carbonaceous grains, therefore it can be photo-processed in the same manner and contribute to chemistry. However, we do not consider this type of grains due to uncertainty in application of the experimental data on the HAC filaments. The mass of a-C mantle on the silicate grains is 20\% from the whole carbonaceous mass locked in dust in the model of \cite{jones13}. Thus, our modelled contribution of hydrocarbons desorbed from dust grains is approximately 20\% smaller than it could be if we considered this type of grains.

We assume that HAC grains are hydrogenated initially so that the fraction of hydrogen atoms ($X_{\rm H}$) equals to 0.62, which is the maximum hydrogen fraction ($X^{\rm H,max}$) according to the model of \cite{jones13}. This state corresponds to the band gap energy equalled to 2.67~eV which is believed to be extremely amorphous structure in the ISM. We limit the fraction of hydrogen atoms in the carbonaceous mantle by this maximum value. We adopt that the minimum $X^{\rm H, min}=0$. The hydrogen mass range is divided into $N_{\rm mH}=5$ bins. The hydrogen mass bin borders can be found through $X^{\rm H}$ via expression
\begin{equation}
m_{\rm bH}^{ij} = j m_{\rm b}^i\left(\frac{X^{\rm H,max}-X^{\rm H,min}}{N_{\rm mH}+1}\right).
\end{equation}

We also assume that large HAC grains cannot be processed deeply inside and set that procession can occur in the surface layer of thickness 200~\AA{} as it was suggested by \cite{jones12_3}. Only the layer can change its hydrogenation level, while the core remains aliphatic. Therefore, for the grains larger than 200~\AA{} all related parameters ($m_{\rm bH}^{ij}$, $X_{\rm H}$, etc.) are calculated for the surface layer.

\section{Results}

\begin{figure*}
	\includegraphics[width=0.45\textwidth]{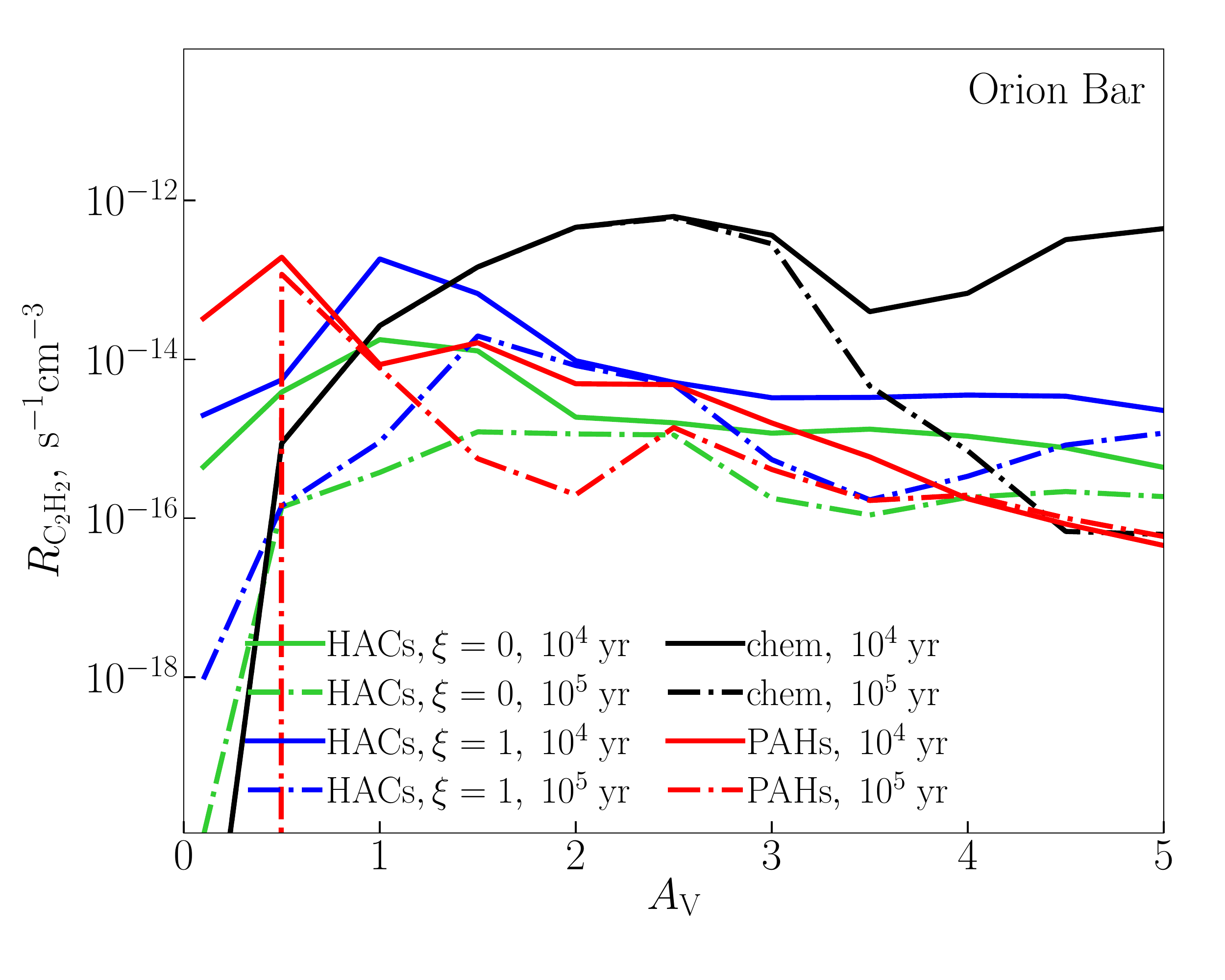}
	\includegraphics[width=0.45\textwidth]{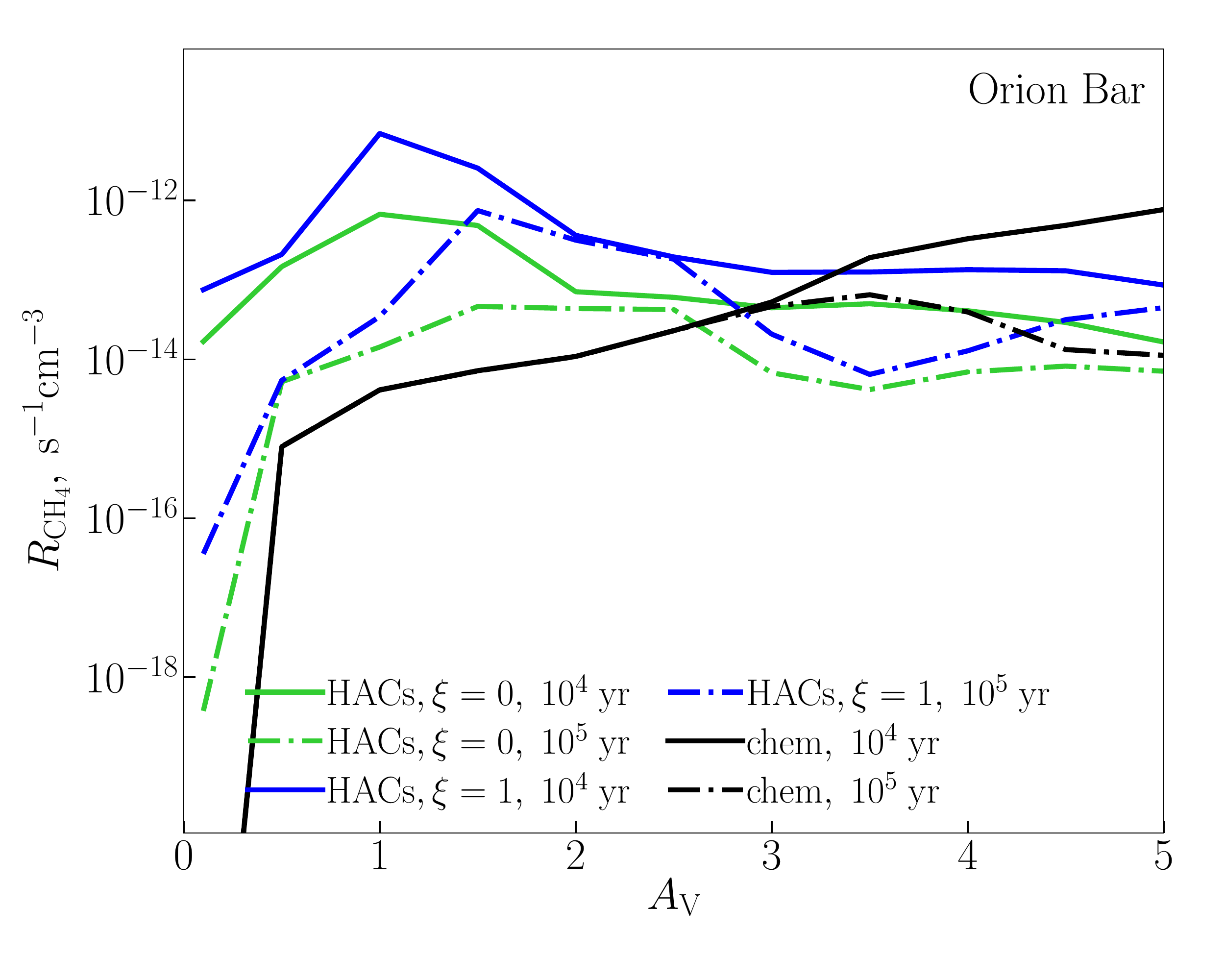}
	\caption{Rates of acetylene (on the left) and methane (on the right) production via the PAH dissociation (red color), the HAC destruction with $\xi=0$ (green color) and  $\xi=1$ (blue color), and in gas phase chemical reactions (black color) in the Orion Bar.}
	\label{compar_orion}
\end{figure*}

\begin{figure*}
	\includegraphics[width=0.45\textwidth]{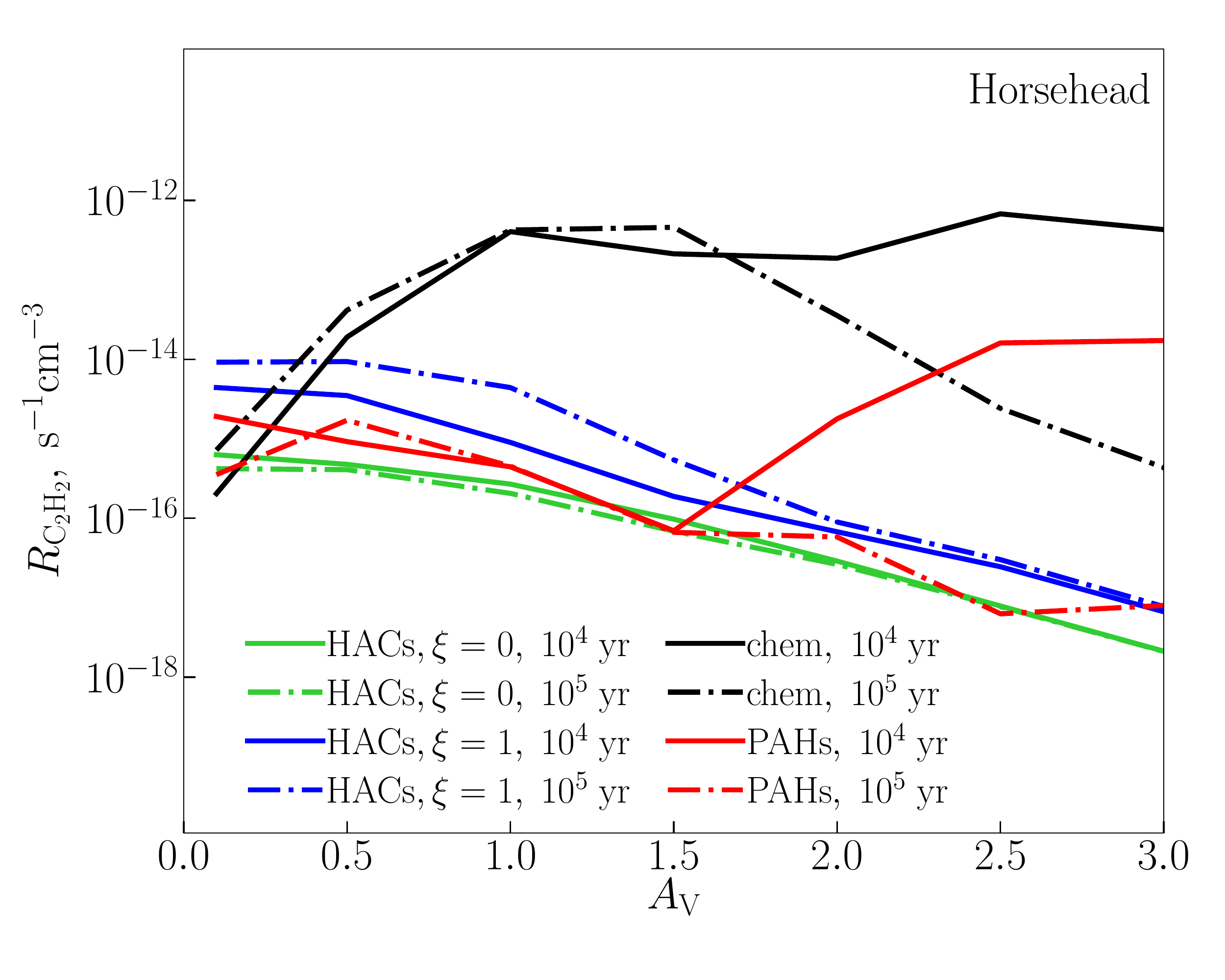}
	\includegraphics[width=0.45\textwidth]{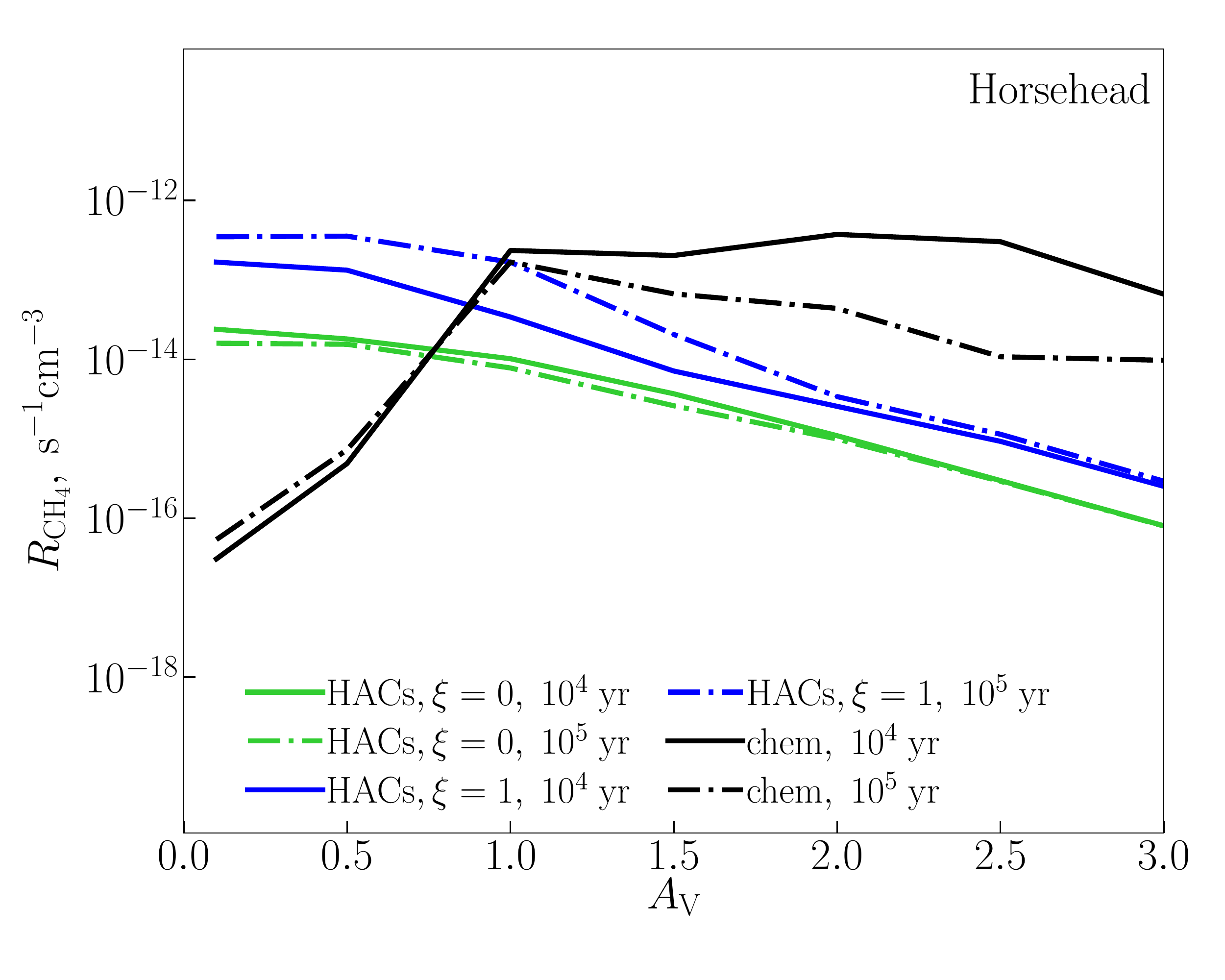}
	\caption{The same as in Fig.~\ref{compar_orion} but for the Horsehead nebula.}
	\label{compar_horse}
\end{figure*}

In Fig.~\ref{compar_orion} we demonstrate the rates of acetylene and methane production due to the PAH dissociation (only acetylene), the HAC destruction with the efficiency of carbon and hydrogen incorporation into a mantle $\xi=0$ and $\xi=1$, and in gas phase chemical reactions in the Orion Bar for 10$^4$ and 10$^5$~yr. The rates of production of acetylene molecules due to the PAH dissociation were taken from our previous work of \cite{murga20_acet}. The rate of acetylene production due to the PAH dissociation is higher than the same rate due to the HAC destruction at $A_{\rm V}<1$ excluding $A_{\rm V}=0.1$ at 10$^5$~yr when PAHs stop producing acetylene. At other locations the PAH rate is mostly smaller than the HAC rate with both $\xi=0$ and $\xi=1$ although the PAH rate is somewhat higher than the HAC rate with $\xi=0$ (namely, at 10$^4$~yr at $A_{\rm V}$ from 1.5 to 3 and at 10$^5$~yr at $A_{\rm V}$ from 2.5 to 4). The rate of acetylene production due to the HAC destruction with $\xi=1$ is higher than the rates with $\xi=0$. The rate of acetylene production due to the HAC destruction dominates over gas phase reactions at $A_{\rm V}\lesssim1.5$ with $\xi=1$ and at $A_{\rm V}\lesssim1.0$ with $\xi=0$ at 10$^4$~yr. At 10$^5$~yr the HAC grains dominate over gas phase reactions in production of acetylene at $A_{\rm V}\approx0.1$ and at $A_{\rm V}>4$ with both $\xi$. Methane production due to the HAC destruction with both $\xi$ is more efficient than the production in gas phase chemical reactions at $A_{\rm V}\lesssim3$ both at 10$^4$ and 10$^5$~yr. The rate of methane production is more efficient due to the HAC destruction when $\xi=1$ as well as for acetylene. 

In Fig.~\ref{compar_horse} we compare the same rates as above but for the Horsehead nebula. The rate of acetylene production due to the HAC destruction with $\xi=1$ are higher than the same rate due to the PAH dissociation for $A_{\rm V}\lesssim1.5$ both at 10$^4$ and 10$^5$~yr, but they are smaller for $A_{\rm V}>1.5$ at 10$^4$~yr. The rates of acetylene production due to the HAC destruction with $\xi=0$ are approximately the same as the rates of production due to PAH dissociation for $A_{\rm V}<1.5$ at 10$^5$~yr, but they are smaller for $A_{\rm V}>1.5$ at 10$^4$~yr. The rates of acetylene production due to the HAC destruction are higher when the parameter $\xi=1$. Gas phase chemical reactions have higher rates of the acetylene production than the HAC destruction at 10$^4$ and 10$^5$~yr over the considered range of $A_{\rm V}$ but for $A_{\rm V}=0.1$ where the HAC rates are either higher or comparable. Methane is produced in gas phase chemical reactions more efficiently than due to the HAC destruction (both $\xi$) at $A_{\rm V}>1$. Again, HAC grains produce methane with higher rates if $\xi=1$. 

\begin{figure*}
	\includegraphics[width=0.45\textwidth]{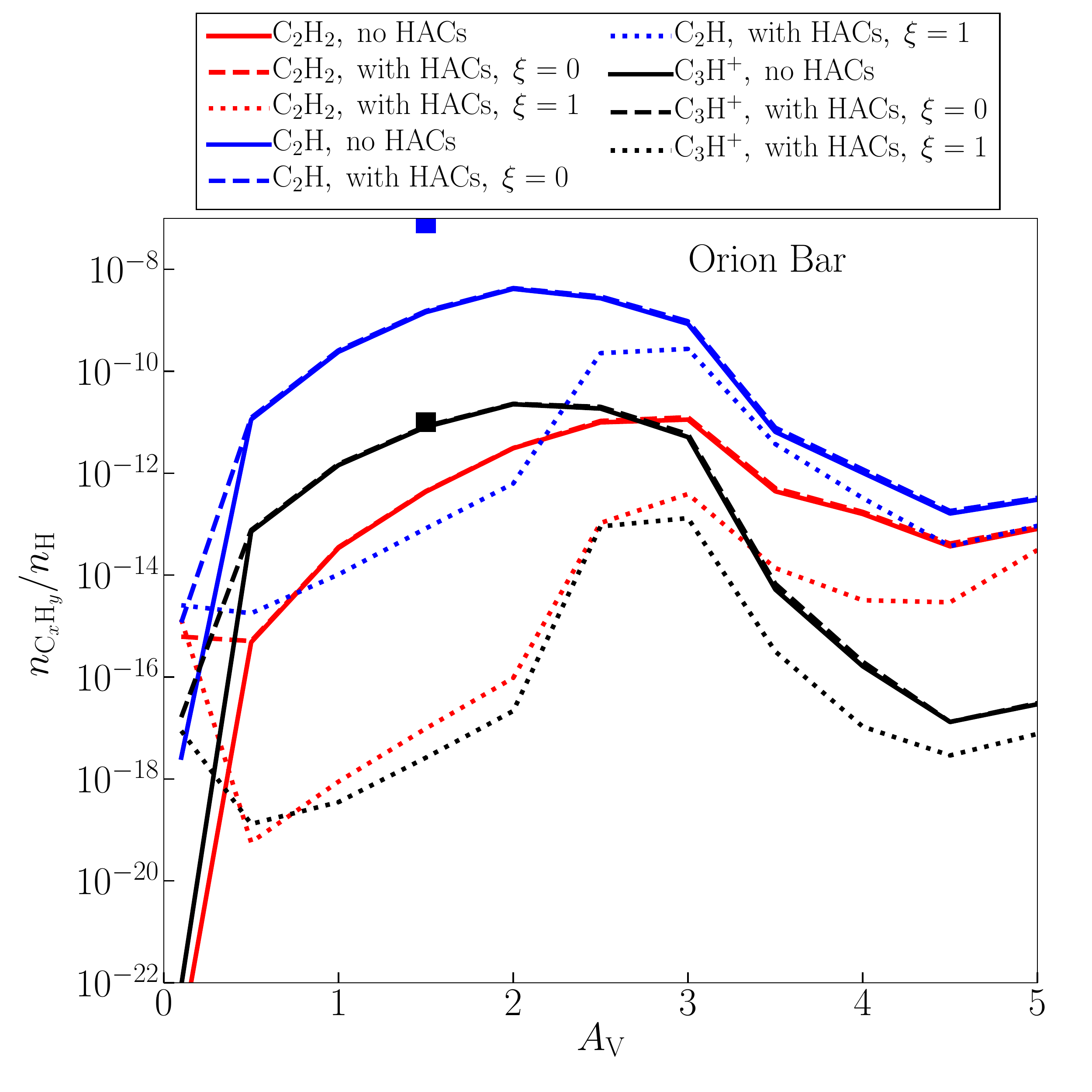}
	\includegraphics[width=0.45\textwidth]{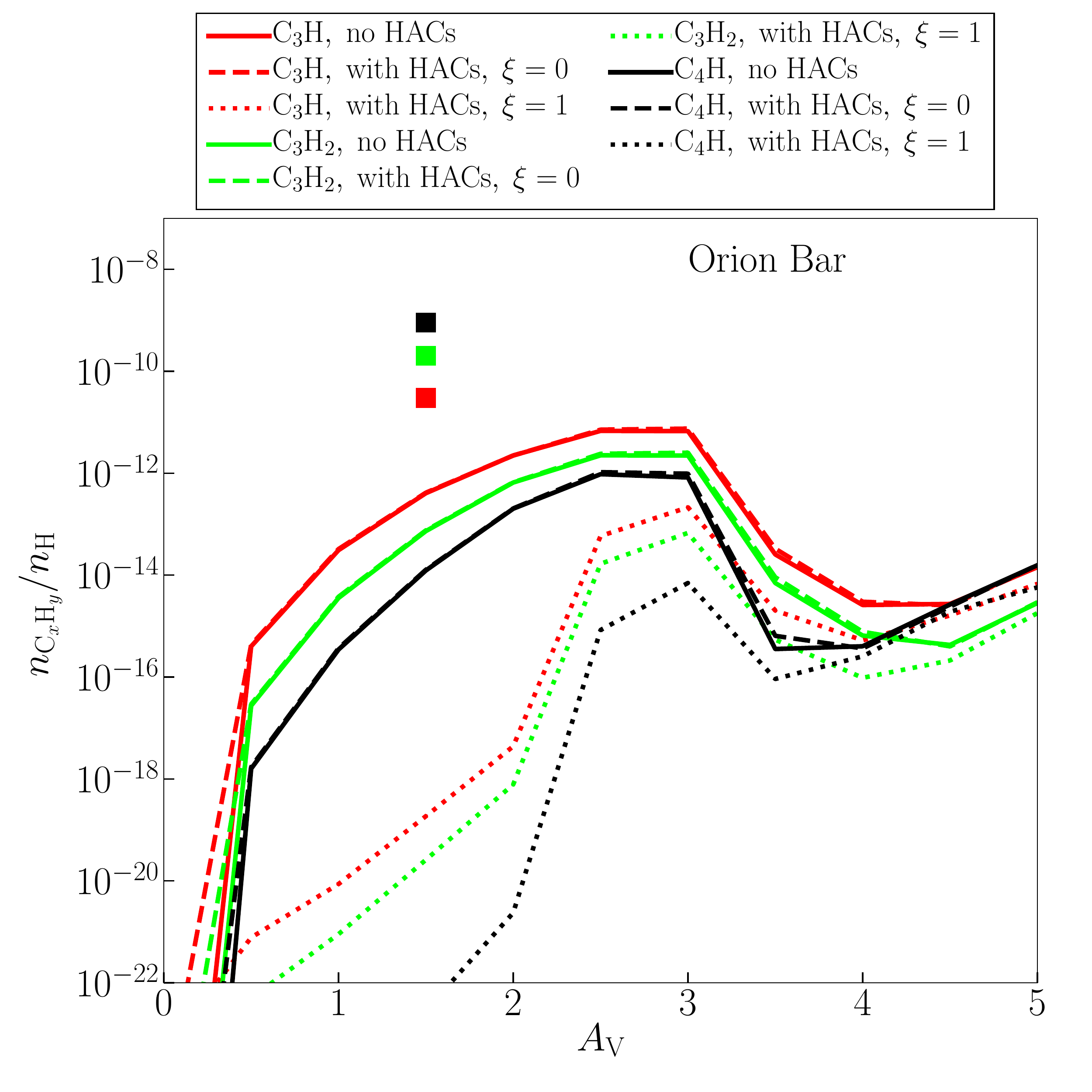}
	\caption{Abundance of small hydrocarbons produced in the gas phase chemical reactions without the evolution of HAC grains (solid lines) and with HAC grains (dashed lines are for $\xi=0$, dotted lines are for $\xi=1$) in the Orion Bar. The observational abundances from Cuadrado et al. (2015) are plotted by squares.}
	\label{ncxhy_av_orion}
\end{figure*}

In Fig.~\ref{ncxhy_av_orion} we present abundances of small hydrocarbons in the Orion Bar calculated with the MONACO model without consideration of the HAC evolution and with their evolution with the accretion parameter $\xi=0$ and $\xi=1$. If the accretion process is suppressed (i.e. $\xi=0$) the contribution of fragments produced due to the HAC destruction increases the modelled abundances for 3-7 orders at $A_{\rm V}=0.1$ where the HAC destruction produces hydrocarbons more efficiently than gas phase chemical reactions. At other locations the contribution gives an insignificant increase. When the accretion process is on, i.e. $\xi=1$, the abundances of small hydrocarbons obtained with the model that takes into account the HAC evolution become lower over considered range of $A_{\rm V}$ but the point $A_{\rm V}=0.1$ where the abundances are the same as obtained with $\xi=0$. The observational abundances from \cite{cuadrado15} are given at $A_{\rm V}=1.5$. Our modelled abundance without HAC grains and with HAC grains are lower for C$_2$H, C$_3$H, C$_3$H$_2$, and C$_4$H but they are consistent for C$_3$H$^{+}$ (without HAC evolution or with it with $\xi=0$).

\begin{figure*}
	\includegraphics[width=0.45\textwidth]{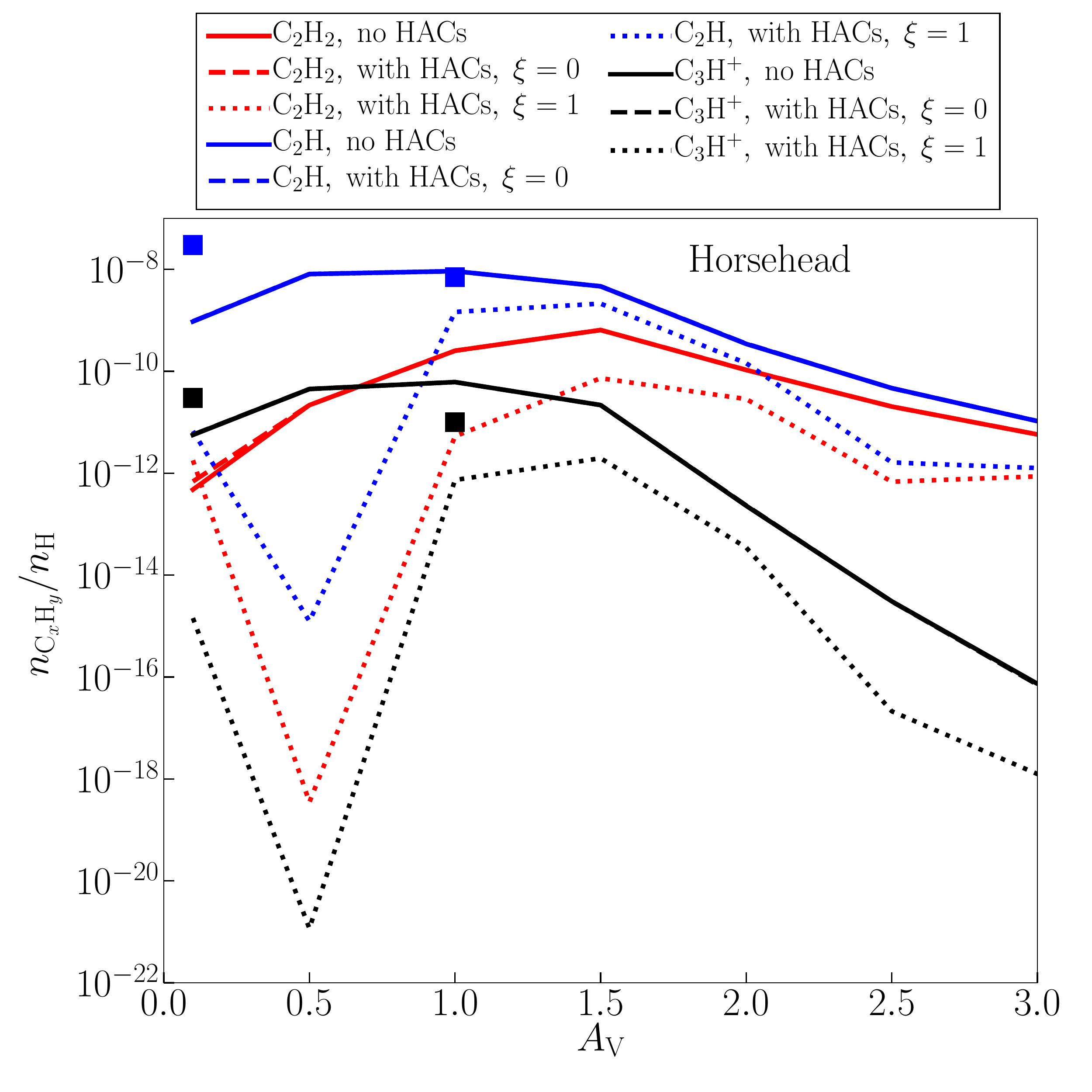}
	\includegraphics[width=0.45\textwidth]{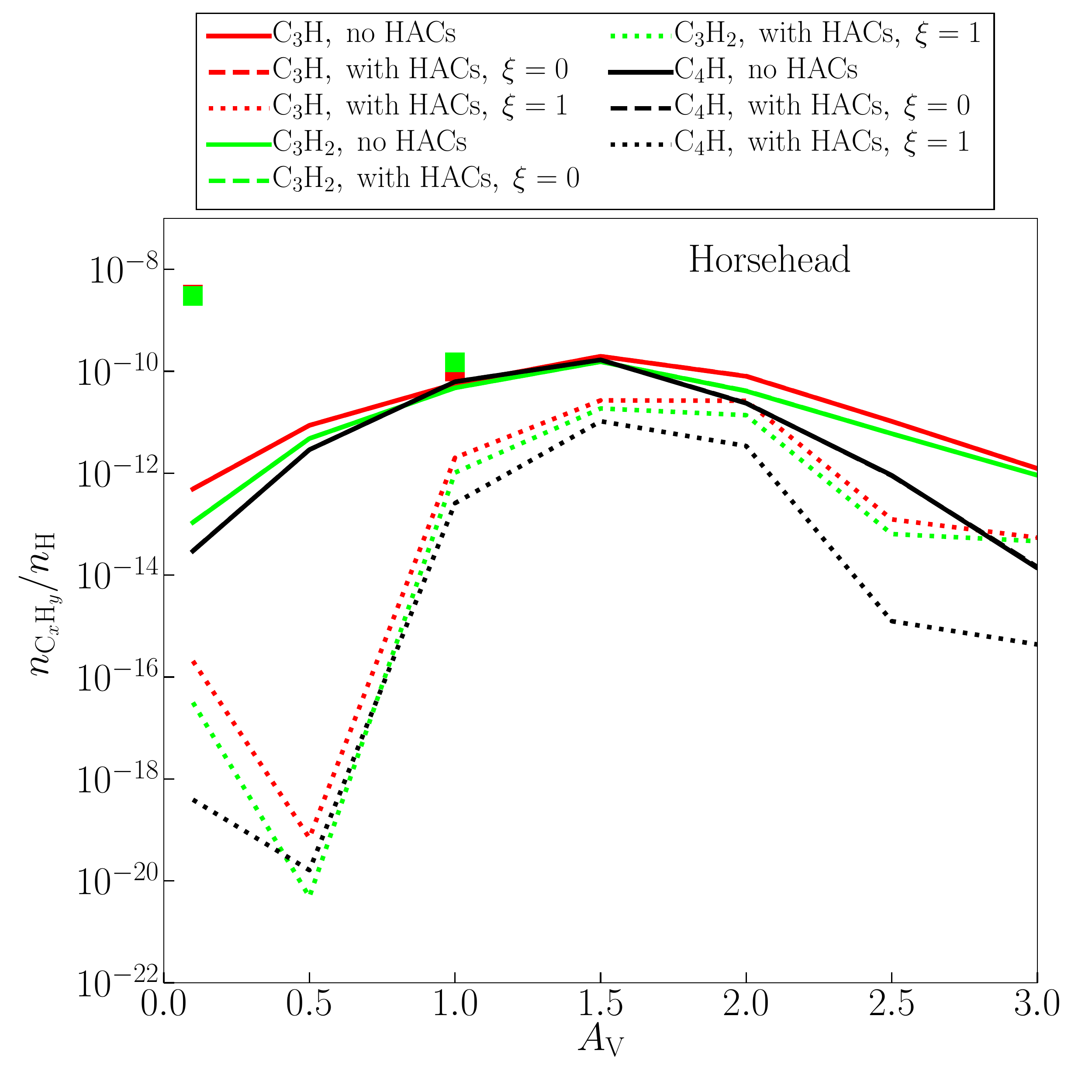}
	\caption{The same as in Fig.~\ref{ncxhy_av_orion} for the Horsehead nebula. The observational abundances from Guzman et al. (2015) are plotted by squares.}
	\label{ncxhy_av_horse}
\end{figure*}

We present the abundances of small hydrocarbons in the Horsehead nebula in Fig.~\ref{ncxhy_av_horse}. It is seen that the modelled abundances without and with the input from the HAC evolution with $\xi=0$ coincide. However, again, if we `switch on' the mantle formation process (i.e. $\xi=1$) the modelled abundances may decrease by many orders of magnitude. Thus, inclusion of the accretion and corresponding formation of the carbonaceous mantle on dust surface reduces the formation of small hydrocarbons. The observed abundances taken from \cite{guzman15} are plotted in Fig.~\ref{ncxhy_av_horse}. The modelled abundances without inclusion of the HAC evolution and with its inclusion with $\xi=0$ almost match the observed values at $A_{\rm}=1.0$ for all species, while the abundance of C$_3$H$^{+}$ is even overestimated. However, at $A_{\rm}=0.1$ the model underestimates the abundances by more than 1-4 orders of magnitude. 

\begin{figure*}
	\includegraphics[width=0.45\textwidth]{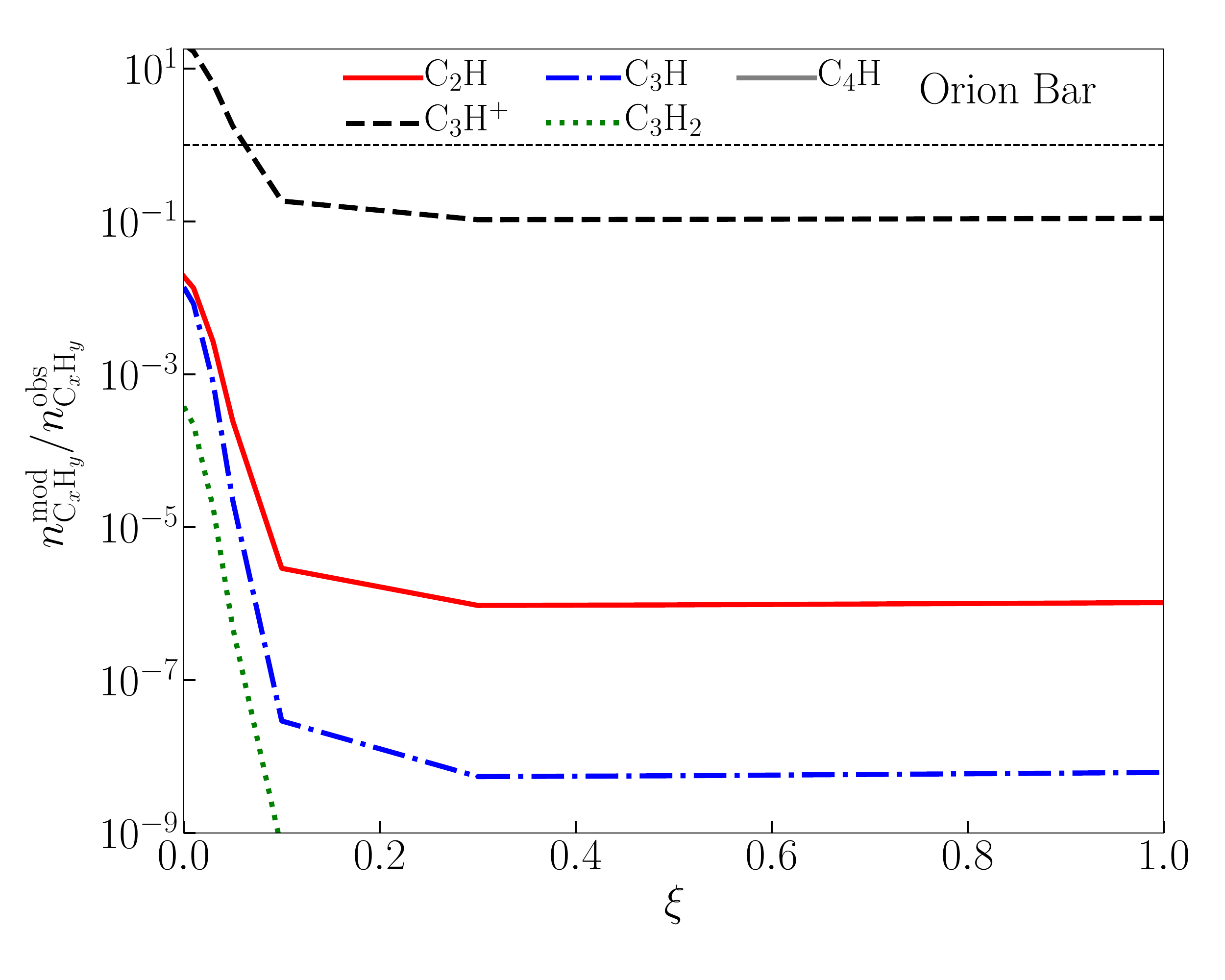}
	\includegraphics[width=0.45\textwidth]{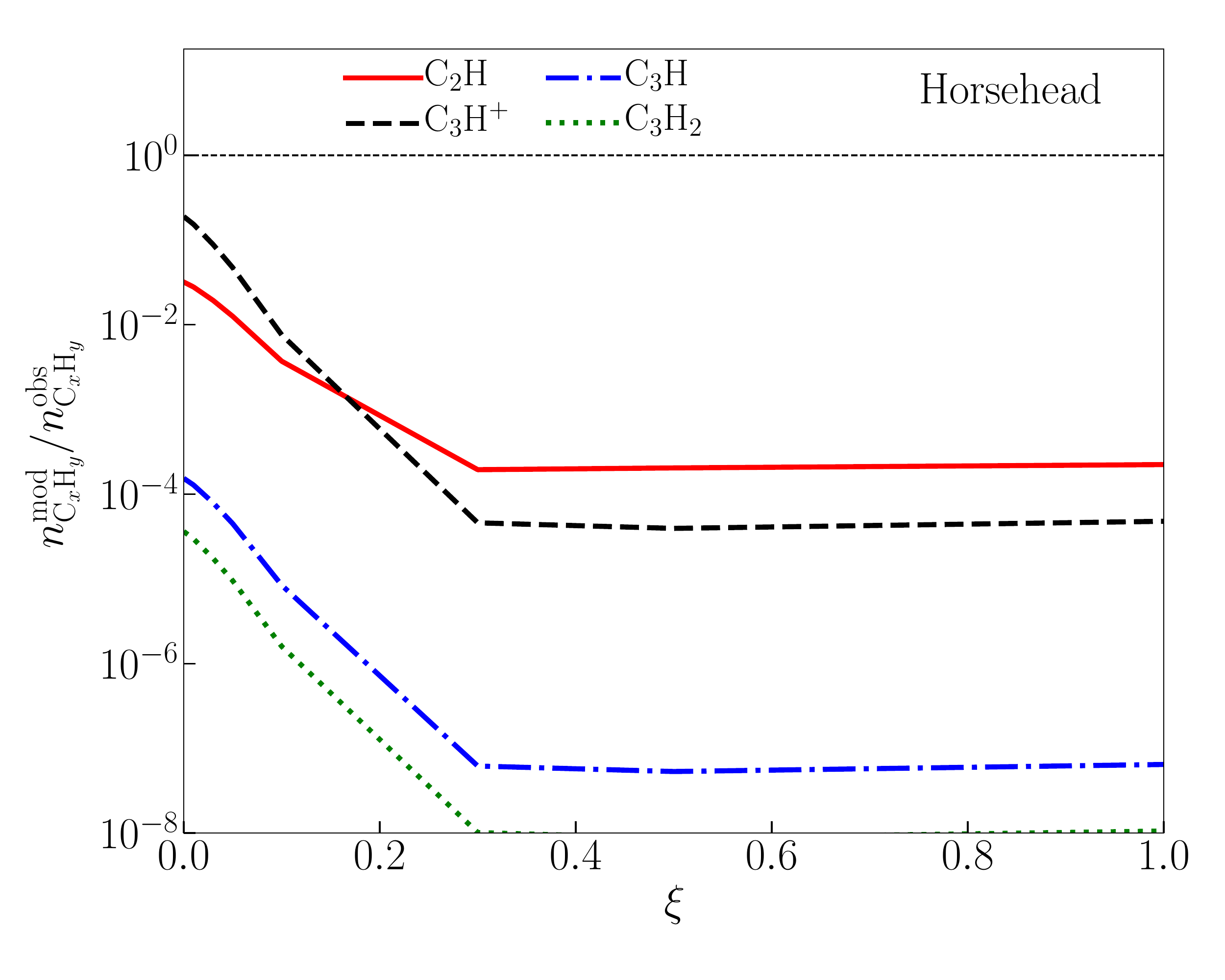}
	\caption{Dependence of the ratio between modelled and observed abundances of small hydrocarbons on the parameter of the accretion efficiency in the Orion Bar at $A_{\rm V}=1.5$ (on the left) and in the Horsehead nebula at $A_{\rm V}=0.1$ (on the right).}
	\label{abund_alpha}
\end{figure*}

The results demonstrated in Fig.~\ref{compar_orion}-\ref{ncxhy_av_horse} indicate the importance of the parameter $\xi$, i.e. the efficiency of carbonaceous mantle formation through accretion of carbon and hydrogen atoms. It was mentioned above that the parameter is undetermined indeed. Therefore we performed the calculations varying the parameter from 0 to 1. We show how the results change with $\xi$ in Fig.~\ref{abund_alpha} presenting the dependence of the ratio between modelled and observed abundances of species  on the parameter $\xi$ at $A_{\rm V}=1.5$ for the Orion Bar and $A_{\rm V}=0.1$ for the Horsehead nebula. As $\xi$ increases, i.e. the processes of accretion and mantle formation start working, the modelled abundances sharply drop by several orders of magnitude and then become almost flat for $\xi>0.3$ for both PDRs.
  
\section{Discussion}

The results of our previous work showed that taking into account of the PAH evolution does not lead to an increase of modelled abundances of small hydrocarbons up to the observational values in the Orion Bar and Horsehead nebula. However, the results of the laboratory experiments on the HAC photo-destruction were promising~\citep{alata14, alata15}, and we attempted to perform similar calculations but taking into account of the HAC evolution in this work. It is found that the modelled abundances are still lower than the observations. The production rates of small hydrocarbons due to the HAC destruction are still lower than it is needed. 

The rates of production of small hydrocarbons can be varied within 1-2 orders depending on the adopted value of the parameter $\xi$ which controls the efficiency of the accretion and formation of the carbonaceous mantle. As $\xi$ increases, the rates of production of destruction products increase too. But along with it the final abundances of small hydrocarbons become even lower. It is caused by the depletion of carbon atoms from the gas phase due to their accretion onto dust grains. The carbonaceous amorphous  mantle is formed on the surface, and this mantle is dissociated due to photo- and thermo-desorption processes. As a result, carbon locked in various hydrocarbons participates in chemical reactions. According to our results the observed hydrocarbons are formed more efficiently if C and C$^{+}$ participate in chemical reactions in their atomic form instead of being locked in some hydrocarbons. Only at $A_{\rm V}=0.1$ in the Orion Bar the abundances of observed hydrocarbons obtained with HAC contribution with both $\xi$ are higher than the abundances obtained without HAC contribution. At this point radiation field is more intense than at other locations, and, apparently, chemical reactions with influent hydrocarbons proceed with rates which are high enough to increase abundances of observed hydrocarbons.

It can be seen in Fig.~\ref{abund_alpha} that the modelled abundances steeply drop and become flat for $\xi>0.3$ for the radial points corresponding to $A_{\rm V}=1.5$ in case of Orion Bar and $A_{\rm V}=0.1$ in case of the Horsehead nebula. This plateau appears because the significant fraction of gaseous C and C$^+$ is either locked in small hydrocarbons  or settled on the dust surface at $\xi \approx0.3$. More efficient mantle formation does not lead to decrease in the efficiency of formation of hydrocarbons because the gaseous carbon is already reduced significantly at $\xi \approx0.3$, and further decrease in the gaseous carbon does not affect the results. Some gaseous carbon comes from dissociation of hydrocarbons in chemical reactions and this carbon again may accrete onto dust surface and participate in different chemical reactions but the amount of the released carbon is not enough to achieve the abundances obtained with the model at $\xi>0.3$. In the dense PDRs, at least one third part of total carbon abundance is in the form of C$^+$ (i.~e. C$^+$/H~$\sim 10^{-4}$) as it was shown by theoretical and observational studies \citep[e.~g. by][]{tielens05, goicoechea15, pabst17, 2020MNRAS.497.2651K} with much less fraction of other C-bearing species. As the dust and gas phases are coupled in our model, the ratio of carbon in these two phases changes with time. Therefore, we can obtain various fractions of carbon locked in HAC grains and available in the gas phase, in accordance with the results of e.~g. \cite{1997ApJ...475..565D, 2012ApJ...760...36P}. Therefore, the process of formation of the carbonaceous mantle should not be efficient under the PDR conditions or, at least, $\xi$ must be lower than 0.3. Nevertheless, it is worth showing that the process may play role in chemistry and may be an inhibitor for the formation of small hydrocarbons. 

\section{Conclusions}

We modelled the evolution of the HAC grains under the conditions of two PDRs, the Orion Bar and the Horsehead nebula. We coupled this model with the gas-grain chemical model MONACO. We calculated abundances of hydrocarbons C$_2$H,  C$_3$H, C$_3$H$^{+}$, C$_3$H$_2$, C$_4$H for which observations in these PDRs are available. We conclude that modelled abundances are lower than the observational values either with or without taking into account the HAC evolution. The contribution of fragments desorbed from the HAC grain surface to chemistry is not enough to increase the modelled abundances up to observational values. Moreover, the model of the HAC evolution is sensitive to the efficiency of the carbonaceous mantle formation. As this process becomes more efficient, the abundances of small hydrocarbons decrease. This anticorrelation is caused by reducing of the gaseous carbon due to accretion. 

\section*{Acknowledgements}

We are grateful to the anonymous referee for insightful comments which helped to improve the paper. We thank Javier Goicoechea who provided the profiles of physical parameters for the Orion Bar PDR. 

The work is supported by the grant of the Russian Science Foundation (project 18-12-00351).

\section*{Data availability}

The data underlying this article are available in Figshare at 
{\tt\url{https://doi.org/10.6084/m9.figshare.20024300}}

\bibliographystyle{mnras} 
\bibliography{refs_hac}

\bsp	
\label{lastpage}
\end{document}